\documentclass[twocolumn,secnumarabic,amssymb, nobibnotes, aps, prl,superscriptaddress]{revtex4-2}

\usepackage{amsmath}
\usepackage{graphicx}
\usepackage{dcolumn}
\usepackage{bm}
\usepackage{tabularx}
\usepackage{upgreek}
\usepackage{eufrak} 
\usepackage{dcolumn}
\usepackage{bm}
\usepackage{siunitx}
\usepackage{dcolumn}
\usepackage{microtype}

\newcommand*{\balancecolsandclearpage}{
  \close@column@grid
  \clearpage
  \twocolumngrid
}

\usepackage{xfp}
\usepackage{chngcntr}

\begin{document}

\author{Marlo Kunzner}\email{marlo.kunzner@dlr.de}\affiliation{German Aerospace Center (DLR) Institute of Frontier Materials on Earth and in Space Functional, Granular, and Composite Materials 51170 Cologne, Germany}

\author{W. Till Kranz}
\affiliation{German Aerospace Center (DLR) Institute of Frontier Materials on Earth and in Space Functional, Granular, and Composite Materials 51170 Cologne, Germany}

\author{Matthias Sperl}\email{Matthias.Sperl@dlr.de}\affiliation{German Aerospace Center (DLR) Institute of Frontier Materials on Earth and in Space Functional, Granular, and Composite Materials 51170 Cologne, Germany}\affiliation{Department for Theoretical Physics, University of Cologne, Germany}

\author{Jan Philipp Gabriel}\email{Jan.Gabriel@dlr.de}\affiliation{German Aerospace Center (DLR) Institute of Frontier Materials on Earth and in Space Functional, Granular, and Composite Materials 51170 Cologne, Germany}

\title{Time–state superposition in non-equilibrium fluidized granular matter}

\begin{abstract}
Despite being intrinsically athermal and strongly driven, granular materials can exhibit remarkably glass-like dynamics. Whether their rheology can be described by the same scaling concepts remains an open question. Here, we investigate the linear viscoelastic response of an air-fluidized granular bed using small-amplitude oscillatory shear over a broad range of fluidization states. We show that the frequency-dependent spectra collapse onto a single master curve when shifted by a state-dependent relaxation time, establishing a time-state superposition principle analogous to time-temperature superposition in molecular glasses. The master curve spans more than five decades in relaxation time and is quantitatively described by a Cole-Davidson relaxation spectrum. By comparison with continuous shear measurements, we identify tribocharging as the origin of history-dependent deviations from universal scaling. Our results demonstrate that fluidization primarily rescales a single structural relaxation time while preserving the underlying relaxation spectrum, establishing a direct connection between the rheology of driven granular matter and molecular glass-forming liquids.
\end{abstract}

\maketitle

Time-temperature superposition (TTS) is among the most successful concepts in the rheology of molecular liquids and glasses. It allows viscoelastic spectra measured at different temperatures to be shifted onto a single master curve \cite{olsen2001time,eliasen2025high,mikkelsen2022dielectric}. The existence of such a scaling principle implies that the underlying relaxation spectrum remains unchanged while only a single characteristic relaxation time varies with the control parameter. 

\begin{figure}[t!]
\centering
\includegraphics[trim={0 0.5cm 0 0 },clip,width=0.99\linewidth]{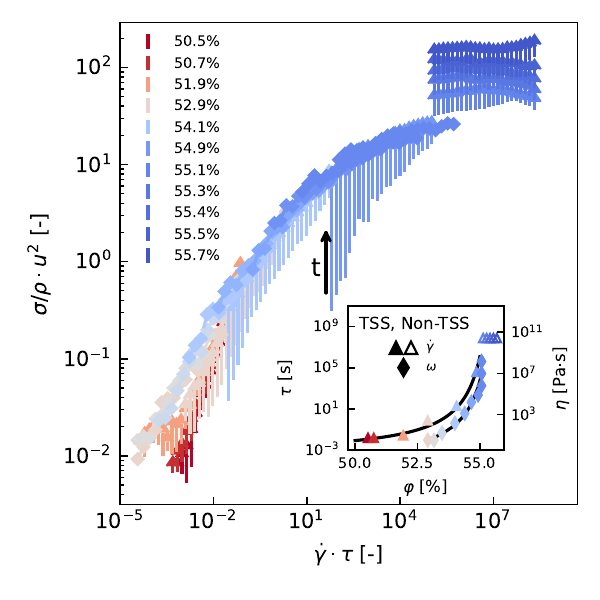}
\caption{
Dimensionless shear stress $\sigma/\rho u^2$ as a function of dimensionless shear rate $\dot\gamma\tau$ for different volume fractions (color coded); symbols denote the stress amplitude of small-angle oscillatory shear measurements, while lines denote continuous shear measurements and an evolution in time. Note that the data approximately collapse onto a master curve except for the highest packing fractions, indicating a time-state superposition principle (TSS) (see main text for details). Inset: The characteristic time scale $\tau$ facilitating the data collapse (left axis) or viscosity $\eta$ (right axis) as a function of packing fraction $\varphi$. Open (closed) symbols denote data where TSS holds (fails). The solid line is a fit to the Vogel-Fulcher law, Eq.~(\ref{eqVF}).
}
\label{fig1a}
\end{figure}

\begin{figure}[t!]
\centering
\includegraphics[trim={0 0.3cm 0 0 },clip,width=1\linewidth]{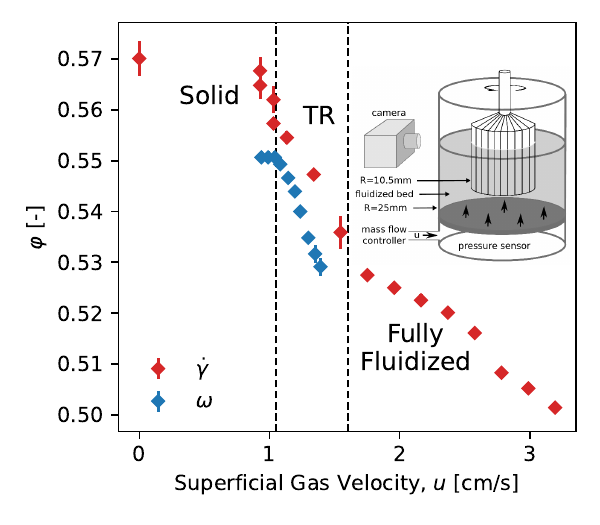}
\caption{
Volume fraction $\varphi$ as a function of fluidization speed $u$ measured during the continuous shear (red) and small-angle oscillatory shear (blue) measurements. The vertical lines denote the boundaries of effectively solid and fully fluidized behavior with a partially fluidized, transition (TR) regime in between. Inset: Fluidized-bed setup (see main text for details).
}
\label{fig1b}
\end{figure}

Driven granular materials are dissipative systems whose dynamics are governed by friction, inelastic collisions, and external forcing rather than by thermal fluctuations. Nevertheless, dense granular systems often exhibit slow relaxation, dynamical heterogeneity, and glass-like behavior that closely resemble phenomena observed in molecular materials \cite{Reis2007,abate2006approach,dauchot2005dynamical,dangelo2025rheological,kunzner2025systematics}. This raises the question of whether concepts originally developed for thermal equilibrium systems can also describe the dynamics of driven granular matter \cite{menon1997particle,kranz2010glass,kawasaki2015diverging,d2023manifold}.

Air-fluidized granular beds provide an ideal platform to address this question. In these non-equilibrium systems, an upward gas flow suspends the particles and injects energy, resulting in fluid-like behavior despite the dissipative particle interactions. Such systems are widely used in industrial processes, including coating, mixing, agglomeration, and particle-size reduction \cite{baerns1966effect,vaidheeswaran2022validation,choi1985dynamic}. The degree of fluidization can be continuously controlled through the gas flow rate, allowing systematic variation of the particle dynamics and packing structure.

Here, we show that the linear viscoelastic response of air-fluidized granular matter obeys a time-state superposition (TSS) principle analogous to TTS (Fig.~\ref{fig1a}). Using low-amplitude oscillatory rheology, we measure the complex shear modulus over a broad range of fluidization states while minimizing perturbations to the granular structure. By comparing oscillatory and steady-shear measurements, we further identify systematic modifications caused by continuous agitation, including packing-fraction changes and tribocharging. We show that the frequency-dependent spectra collapse onto a universal master curve when shifted by a state-dependent time scale, establishing a direct analogue of time-temperature superposition in a driven athermal system.

\textit{Experimental Procedure---} Rheological measurements were performed in a Taylor-Couette geometry using a commercial, torque-controlled rheometer (MCR 102, Anton Paar) equipped with a Powder Flow Cell (Fig.~\ref{fig1b}) that has been used before for similar measurements \cite{kunzner2025systematics,dangelo2025rheological,DIN_ISO}. The fluidized bed is comprised of \qty{59.621}{gram} of polystyrene beads of mean diameter \qty{130}{\micro\metre} (bulk density $\rho_\mathrm{b}=\qty{1.05}{\gram\per\cubic\centi\metre})$. The superficial gas velocity $u$ of the fluidizing air was controlled by a flow controller (Bürkert 8712), and the resulting bed height was recorded by a camera (Sandberg 134-39). From the bed height and the rheometer geometry, we obtain the globally averaged volume fraction $\varphi(u)$ (Fig.~\ref{fig1b}) as a function of the fluidization speed $u$ and the mass density $\rho=\varphi\rho_\mathrm{b}$ \cite{kunzner2025systematics}. We investigated fluidization speeds from \qtyrange{0}{3.2}{\centi\metre\per\s}, covering the transition from a solid granular packing to a fully fluidized bed. In contrast to $\varphi(u)$, we do not have access to the granular temperature's (mean kinetic energy per particle) dependence on the fluidization speed $u$. On dimensional grounds, it has to be $\propto\rho u^2$, and we assume the unknown constant of proportionality to be approximately constant within the parameter range under investigation. 

\begin{figure}[t!]
\centering
\includegraphics[trim={0 0.5cm 0 0 },clip,width=0.99\linewidth]{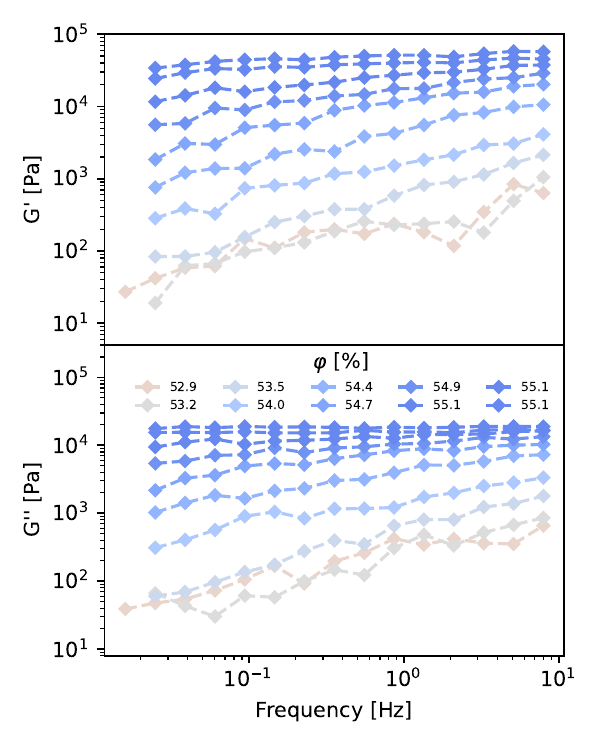}
\caption{
Storage modulus $G'$ (upper panel) and loss modulus $G''$ (lower panel) as a function of frequency $\nu$ measured at different volume fractions $\varphi$ (color coded) as indicated.
}
\label{fig2}
\end{figure}

\begin{figure}[t!]
\centering
\includegraphics[trim={0 0.5cm 0 0 },clip,width=0.99\linewidth]{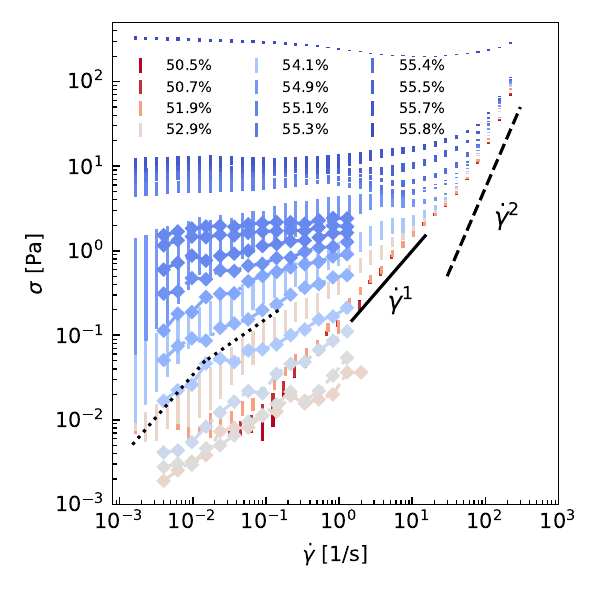}
\caption{
Shear stress $\sigma$ as a function of shear rate $\dot\gamma$ for different packing fractions $\varphi$ (color coded). Solid lines denote continuous shear measurements, while symbols show stresses calculated from the small-angle oscillatory shear measurements using Eq.~(\ref{eq:CM}). The solid and dashed black lines indicate Newtonian and Bagnold behavior, respectively. The dotted line indicates the approximated $\dot\gamma$(52.9\%)=0.015 where the systems change from viscose to Newtonian flow.
}
\label{fig3}
\end{figure}

We performed two types of rheological measurements for each fluidization speed: (i) Small amplitude oscillatory shear (SAOS), where we imposed a sinusoidal time dependent strain $\gamma(t) = \gamma_0\sin(2\pi\nu t)$ with a fixed strain amplitude $\gamma_0 = 4\times10^{-5}$ and frequencies $\nu$ from \qtyrange{0.02}{7.96}{\hertz} on the inner cylinder \cite{eirich2012rheology,Macosko1993Rheology}. The resulting stress signal, $\sigma(t) = \sigma_0(\nu)\sin[2\pi\nu + \delta(\nu)]$, can be presented as the complex, frequency dependent shear modulus $G^*(\nu) = [\sigma_0(\nu)/\gamma_0]\exp[i\delta(\nu)] \equiv G'(\nu) +iG''(\nu)$ (Fig.~\ref{fig2}). Here, every data point is the average over nine independent measurements. (ii) Continuous shear (CS), where we imposed a constant rotation rate $\Omega$ at the inner cylinder and recorded the corresponding stress $\sigma(\Omega)$ at the inner cylinder once a stationary state has been reached (Fig.~\ref{fig3}) and Refs.~\cite{dangelo2025rheological,d2023manifold,kunzner2025systematics} for similar measurements. The mapping between $\Omega$ and the shear rate at the inner cylinder depends on the rheology. For simplicity, here we use an effective shear rate $\dot\gamma = K_\mathrm{N}\Omega$ \cite{dangelo2025rheological}, with the Newtonian strain constant $K_\mathrm{N} = 2.6$, throughout. For a given experiment $(u, \Omega)$, the shear stress $\sigma(\Omega)$ slowly increases in time. An effect which we attribute to collisional tribocharging of the granular particles \cite{haeberle2018double,preud2023tribocharging} and which we represent by the lines in Fig.~\ref{fig3} with the stresses increasing over time from the bottom to the top of a line.

\textit{Results---}Let us begin by stressing that the volume fraction $\varphi$ is not uniquely controlled by the fluidization speed $u$ but also depends on the type of shear experiment (Fig.~\ref{fig1b}). For SAOS (Fig.~\ref{fig2}), we find shear moduli that systematically increase with packing fraction and oscillation frequency. The phenomenology of the CS measurements (Fig.~\ref{fig3}) is more complex \cite{dangelo2025rheological}: At the highest densities, non-monotonicities in the flow curves indicate shear banding \cite{d2023manifold}. At high shear rates, where shear heating dominates over fluidization insofar as the energy input to the system is concerned, the flow curves display Bagnold scaling, $\sigma\sim{\dot\gamma}^2$ \cite{Bagnold1954, andreotti2013granular,shi2020steady}. Once shear heating becomes subdominant at intermediate shear rates, a shear thinning regime, $\sigma \approx $ const., develops that extends to smaller and smaller shear rates, the higher the density. From the lowest horizontal flow curves, we estimate a minimal yield stress on the order of $\sigma_\mathrm{y} = \qty{5}{Pa}$. Finally, at low densities and shear rates we find Newtonian rheology, $\sigma\sim\dot\gamma$. An crossover from Newtonian to shear thinning can be approximated to $\dot\gamma(52.9\%)^1=0.015$, see doted lines in fig \ref{fig3}.

To compare the CS and SAOS measurements, we employ a Cox-Merz-type rule \cite{rathinaraj2022cox}, plotting the stress amplitude 
\begin{equation}
    \sigma(\dot\gamma) \approx \sigma_0(\nu=\dot\gamma/2\pi\gamma^*)
    \label{eq:CM}
\end{equation} 
of the SAOS as a function of an effective shear rate  $\dot\gamma\equiv2\pi\nu\gamma^*$ where $\gamma^*=2.5\times10^{-2}$ is an effective yield strain (see below). For $\varphi=52.9\%$ we estimate the structural relaxation time $\tau(52.9\%)=1/2\pi\nu=\qty{1.7}{s}$ (cf.~Fig.~\ref{fig3}).

Given the empirical nature of the Cox-Merz rule, the results of both CS and SAOS are surprisingly compatible. 

\textit{Time-State Superposition---}For dimensional reasons, we have $\sigma(\dot\gamma\mid u) = \rho u^2\hat\sigma(\dot\gamma\mid \varphi)$ with a dimensionless function $\hat\sigma(\dot\gamma)$. In case $\hat\sigma$ is characterized by a single time scale $\tau = \tau(\varphi)$, there will be a universal function $\check\sigma(\dot\gamma\tau)$ of the dimensionless shear rate $\dot\gamma\tau$ such that $\sigma(\dot\gamma\mid u) = \rho(u)u^2\check\sigma[\dot\gamma\tau(\varphi(u))]$. Assuming Cox-Merz, this can be translated to the SAOS case, i.e., $\sigma_0(\nu\mid u) = \rho(u)u^2\check\sigma[2\pi\gamma^*\nu\tau_\omega(\varphi(u))]$. Note that the timescales are expected to differ between CS and SAOS, $\tau$ and $\tau_\omega$, respectively, as different cohesion and tribocharging states are probed.

\textit{A priori}, we do not know whether the fluidized bed is indeed characterized by a single time scale. However, we can try to collapse our measurements at different densities $\varphi$ by using $\tau(\varphi)$ [$\tau_\omega(\varphi)$, respectively] as a fit parameter and the $\tau(52.9\%)$ from above to fix an overall time scale (appendix \ref{DMP}). As can be seen in Fig.~\ref{fig1a} a convincing data collapse can indeed be achieved. More precisely, the SAOS data can be fully described in terms of $\check\sigma(\nu\tau_\omega)$, a fact that we propose to call \emph{time-state superposition} (TSS). At the highest densities, the resolution is not sufficient to tell the packing fractions apart, though. However, they are expected to smoothly increase with decreasing $u$. The CS data collapse on the same master curve at intermediate densities. At the low density end, deviations become apparent, and in the fully shear-thinning regime at the highest densities, the data collapse clearly fails. This is not surprising as the relaxation time $\tau$ is effectively infinite in this regime and can no longer be the characteristic parameter.

In the relevant range, the relaxation times can be captured by a Vogel-Fulcher law 
\begin{equation}
    \tau(\varphi), \tau_\omega(\varphi) 
    \propto \exp[\Delta\varphi/(\varphi_\mathrm{g} - \varphi)]
    \label{eqVF}
\end{equation}
in terms of the density scale $\Delta\varphi$. The relaxation times diverge towards the critical packing fraction, $\varphi_\mathrm{g}=55.9\%$. Note that the CS system exhibits slower structural relaxation, $\tau<\tau_\omega$, than the SAOS system.   

\begin{figure}[t!]
\centering
\includegraphics[width=0.99\linewidth]{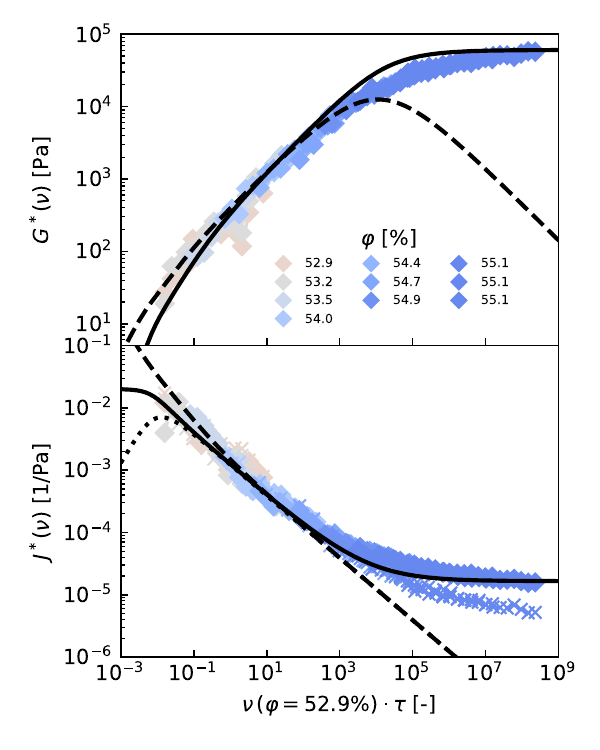}
\caption{
Fit of the compliance-based model, Eq.~(\ref{eq:CD_model}), to the master curve. The frequency dependence is shown in terms of (a) the complex shear modulus and (b) the complex compliance.
}
\label{fig5}
\end{figure}

\textit{A Compliance-based Model.---} Now that we have established that the rheology of the fluidized bed in the Newtonian regime and into the crossover to the shear thinning regime is captured by a single relaxation time $\tau$, we can attempt a quantitative description by a suitable model.  

Following approaches commonly used for molecular glass-forming liquids, we model the compliance as
\begin{equation}
    J^*(\omega) = J_{\infty} + \frac{\Delta J}{i\omega\tau_\mathrm{peak}} + \frac{\Delta J}{\left(1 + i\omega\tau_\mathrm{cd}\right)^{\beta}}.
\label{eq:CD_model}
\end{equation}

Here, $J_{\infty}$ denotes the instantaneous (high-frequency) compliance, the second term captures the Newtonian flow contribution and contains $\tau_\mathrm{peak}=\tau_\mathrm{cd}/\tan[(\pi/(2(1+\beta))]$ \cite{richert2014supercooled} the peak time constant, with characteristic time $\tau_\mathrm{cd}$ of a Cole-Davidson function and the relaxation strength $\Delta J$. The third term represents a viscoelastic relaxation process described by a Cole-Davidson function and stretching exponent $\beta$. Since $\tau_\mathrm{cd}$ is connected to the peak time constant, the $\alpha$-relaxation can be connected with $\eta=\tau_\mathrm{peak}/\Delta J$, the effective viscosity, by assuming the terminal relaxation controls the time when the system flows. 

Values $\beta < 1$ indicate a broadened distribution of relaxation times. While originally developed for molecular systems, this phenomenological model provides a convenient description of the frequency-dependent mechanical response observed here. In molecular systems $\beta \approx 0.5$ is often found \cite{bohmer2025spectral,pabst2021generic}. In normal liquids, one can find similar power laws in rheological compliance and molecular probes, as for example demonstrated in Glycerol \cite{gabriel2020intermolecular}, alcohols \cite{mikkelsen2022dielectric}. In particular, fluids near a glass transition are expected to show broad relaxation peaks.

By fitting the model to our SAOS data (Fig.~\ref{fig5}), we find $J_\infty = \qty{1.65e-5}{\per\pascal}$, $\Delta J = \qty{0.02}{\per\pascal}$, $\beta=0.5$, and $\tau_\mathrm{peak}(\varphi) = \num{7.26}\times\tau(\varphi)$ to capture the rheology well for frequencies $\nu\lesssim1/\tau_\mathrm{peak}$. At higher frequencies, we expect particle effects to take over that are not captured in the continuum model (cf. Appendix \ref{A:Eigen}). The numerical closeness of $\tau$ and $\tau_\mathrm{peak}$ \textit{a posteriori} supports the interpretation of $\tau$ as a structural relaxation time.

The yield stress $\sigma_\mathrm{y} = \gamma_\mathrm{c}/\Delta J$ is related to $\Delta J$ by a Lindemann-like yield strain $\gamma_\mathrm{c}$ \cite{coquand2025granular} which results in a very plausible $\gamma_\mathrm{c} = 0.1$. Via $\eta(\varphi) = \tau_\mathrm{peak}(\varphi)/\Delta J = \qty{150}{\pascal}\times\tau(\varphi)$ we have access to the Newtonian viscosity, $\eta(\varphi)$, of the fluidized bed (cf. inset of Fig.~\ref{fig1a}), following the Vogel-Fulcher law, Eq.~(\ref{eqVF}), as well.  

Given such a broad relaxation spectrum, we can expect a Cox-Merz rule to hold \cite{rathinaraj2022cox} with a renormalized yield strain $\gamma^* = \beta^{1/(1-\beta)}\gamma_\mathrm{c}=2.5\times10^{-2}$ (see above).

\textit{Discussion---} Studying the rheology of an air-fluidized bed from the Newtonian to the crossover to the shear thinning regime, we have established that its rheology is captured by a single, density-dependent relaxation time $\tau(\varphi)$ that diverges towards the critical density $\varphi_\mathrm{g}=0.559$. As a consequence, a time-state superposition holds. Moreover, we showed that the rheology can be quantitatively described by a Cole-Davidson-type model for the complex compliance, Eq.~(\ref{eq:CD_model}), combining viscous flow with a broad relaxation spectrum. In combination with the value of the stretching exponent $\beta = 0.5$, this explains the success of the Cox-Merz rule, Fig.~\ref{fig3}.

In essence, despite the essentially \emph{non-equilibrium} nature of the fluidized bed, comprised of \emph{underdamped}, inertial particles, we have demonstrated that---as along as we stay away from the shear-dominated Bagnold regime---it behaves much like an \emph{overdamped} colloidal suspension in \emph{thermal equilibrium}: (i) Given material properties, the packing fraction $\varphi$ is the only relevant parameter controlling the rheology (considering that $u$ is modifying the material); (ii) The crossover from Newtonian to viscoelastic behavior is controlled by the competition between the shear rate $\dot\gamma$ and the structural relaxation rate $1/\tau$ \cite{dangelo2025rheological}; (iii) The relaxation time $\tau(\varphi)$ diverges in parallel with the viscosity $\eta(\varphi)$ towards a critical density $\varphi_\mathrm{g}$, above which the fluidized bed behaves as a solid body, while the constitutive particles remain dynamic. I.e., $\varphi_\mathrm{g}$ indicates a granular glass transition \cite{kranz2010glass}. Moreover, the incipient glass transition leads to a Cole-Davidson-type broad spectrum known from other glass formers \cite{binder2003glass,hunter2012physics}. Essential differences, though, are: (i) The only control parameter, the fluidization velocity $u$, simultaneously controls the density \emph{and} the granular temperature; two parameters that can be tuned individually in colloidal suspensions; (ii) While colloidal suspensions are naturally thermostatted by the suspending fluid, the granular particles' kinetic energy is decoupled from the fluidizing air which makes it much easier to reach a shear-dominated regime with its Bagnold behavior, unique to granular fluids \cite{andreotti2013granular}.

One striking observation is the large difference between the relaxation times obtained from continuous shear ($\tau$) and oscillatory measurements ($\tau_\omega$), amounting to approximately $\tau \approx 900\,\tau_\omega$, despite both experiments yielding essentially the same dimensionless yield stress $\sigma_\mathrm{y}/\rho u^2$. During continuous shear, progressive tribocharging increases the cohesion between particles and requires a higher fluidization energy to reach the same packing fraction. Consequently, structurally similar packings can exhibit markedly different relaxation dynamics, demonstrating that packing fraction alone is insufficient to fully characterize the state of the fluidized bed.

The estimated cohesive energy density, $c\approx\qty{45}{\joule\per\cubic\metre}$ (Appendix), exceeds the granular kinetic energy density by more than an order of magnitude. Using the estimate $nT\approx\sigma_\mathrm{y}/10$ to $\sigma_\mathrm{y}/5$ \cite{kranz2018rheology}, we obtain $nT\approx\qtyrange{0.5}{1}{\joule\per\cubic\metre}$, implying $c/nT\gg1$. Strong attractive interactions therefore promote long-lived transient particle networks, while fluidization continuously randomizes their configuration. This combination naturally gives rise to a broad spectrum of structural relaxation times resembling that of molecular glass-forming liquids.

The simultaneous observation of Vogel--Fulcher dynamics, Cole--Davidson relaxation, Cox--Merz correspondence, and time--state superposition is remarkable because these concepts are generally regarded as characteristic signatures of molecular glass formers. Our results demonstrate that they can also emerge in a strongly driven, intrinsically athermal granular system.

More generally, the present results suggest that time--state superposition does not rely on thermodynamic equilibrium but can also emerge in driven steady-state systems, where the relevant energy scale is provided by the granular temperature. This establishes a direct connection between the rheology of driven granular matter and that of molecular glass-forming liquids, and suggests that similar scaling principles may apply across a much broader class of non-equilibrium disordered materials.

\begin{acknowledgements}
\textit{Acknowledgements---} This work was supported by the DLR Space Administration with funds provided by the Federal Ministry for Economic Affairs and Climate Action (BMWK) based on a decision of the German Federal Parliament under grant number 50WM1945 (SoMaDy2).   
\end{acknowledgements}

\bibliography{Rheobib}

%\balancecolsandclearpage
\clearpage

\appendix
\section*{Appendix: Additional Information}

\subsection{Pressure Drop}

\begin{figure}[h!]
\centering
\includegraphics[width=0.99\linewidth]{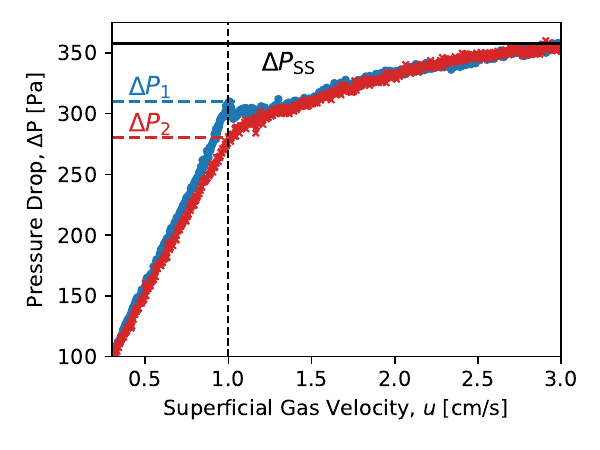}
\caption{
Typical pressure-drop curve for the polystyrene particles ($140 \mu m$) used in this study, shown as the pressure difference $\Delta P$ versus the superficial gas velocity $u$.
}
\label{fig_a}
\end{figure}

Details are given in Kunzner et al. \cite{kunzner2025systematics}. Fig. \ref{fig_a} shows a pressure drop measurement, with the dashed line indicating where the system is fluidized and the solid line indicating the pressure plateau expected from the sample weight. 
\begin{equation}
    \label{eq:Pressure}
    \Delta P = m\cdot g /(\pi\cdot(R_0^2-R_i^2))
\end{equation}\cite{Gottschalk1986,dangelo2025rheological}
Here $\Delta P$ is the pressure difference between filled and empty cell, $m$ the mass, $g$ gravitational acceleration, and $R_0,R_i$ the outer and inner radius, respectively. The cohesive energy density can be estimated from the pressure overshoot $c = |\Delta P_1 - \Delta P_\mathrm{SS}| \simeq\qty{45}{\joule\per\cubic\metre}$ \cite{affleck2023novel,hsu2018analysis}.

\subsection{Measurement procedures}

\begin{figure}[t!]
\centering
\includegraphics[width=1\linewidth]{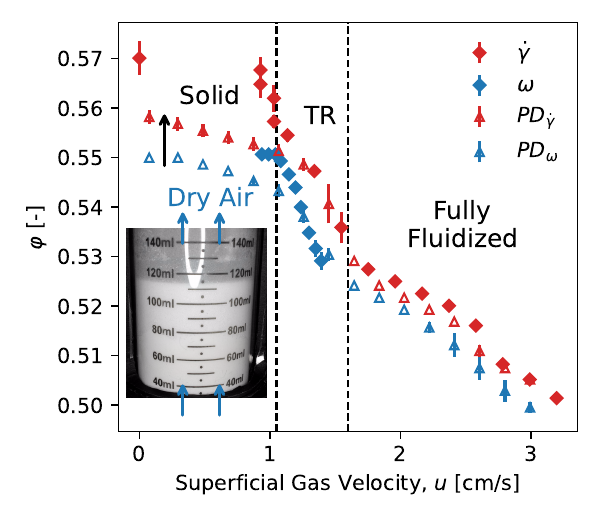}
\caption{Volume fraction $\varphi$ as a function of fluidization speed $u$ measured during the continuous shear (red) and small-angle oscillatory shear (blue) measurements. The vertical lines denote the boundaries of effectively solid and fully fluidized behavior with a partially fluidized, transition (TR) regime in between.
}
\label{figb}
\end{figure}

To establish the relevant fluidization states and assess possible protocol-dependent effects, we first compare the packing fractions obtained using the different measurement procedures. Figure~\ref{figb} summarizes the global packing fraction $\varphi(u)$ measured during oscillatory rheology, steady rotational shear, and pressure-drop experiments performed between the rheological measurements. Filled symbols denote packing fractions measured during rheological experiments under either continuous rotational shear at a constant shear rate $\dot{\gamma}$ or oscillatory shear at angular frequency $\omega$. Open symbols correspond to pressure-drop measurements (PD$_{\dot{\gamma}}$ and PD$_{\omega}$), which were recorded between the rheological measurements. For these measurements, the gas flow rate was ramped from 0 to 3~L$\cdot$min $^{-1}$ or 3~cm/s in order to characterize the fluidization behavior and prepare the system for the subsequent measurement sweep.

A systematic shift in the measured packing fractions is observed between the rotational and oscillatory protocols. Over a broad range of flow rates, the states probed during rotational shear exhibit higher packing fractions than those obtained during oscillatory measurements. The difference is particularly pronounced near the onset of fluidization, where small changes in $u$ lead to large variations in $\varphi$, reflecting the high sensitivity of the system in this regime.

We attribute the increased density observed during rotational measurements to the disruption of bubbles and the resulting reorganization of the granular packing under continuous shear. In contrast, the deviation between rotational and oscillatory measurements at low flow rates is likely associated with electrostatic charging. Continuous particle–particle and particle–wall contacts during rotation promote charge accumulation, thereby enhancing attractive interparticle interactions and stabilizing denser packing configurations. This interpretation is consistent with the substantially longer duration of the continuous-shear protocol, which comprises five repetitions over approximately seven days, compared to nine SAOS repetitions completed within about 2.75 days. These observations demonstrate that the measured packing fraction is not solely determined by the imposed gas flow rate but also depends on the mechanical history and preparation protocol of the granular system.

\subsection{Characteristic Eigenfrequency}\label{A:Eigen}

We estimate the characteristic Eigenfrequency of individual particles. Using a Young's modulus $E = \qty{3}{\giga\pascal}$, the longitudinal sound velocity is given by $v_L \approx \sqrt{E/\rho_\mathrm{b}}$. This yields $v_L \approx \qty{1700}{\metre\per\s}$. Following classical elasticity theory \cite{lamb1881vibrations,love2013treatise}, the fundamental vibrational frequency of a spherical particle can be estimated as
\begin{equation}
f = \frac{\alpha v_L}{2\pi R}.
\end{equation}
With $\alpha \approx 2.2$, this results in $f\approx\qty{9}{\mega\hertz}$. This frequency is several orders of magnitude higher than the experimental range, supporting the interpretation that the observed high-frequency deviations are not governed by particle eigenmodes but rather by the breakdown of the collective viscoelastic description.

\subsection{Dimensionless Master Plot}
\label{DMP}
\begin{figure}[h!]
\centering
\includegraphics[width=1\linewidth]{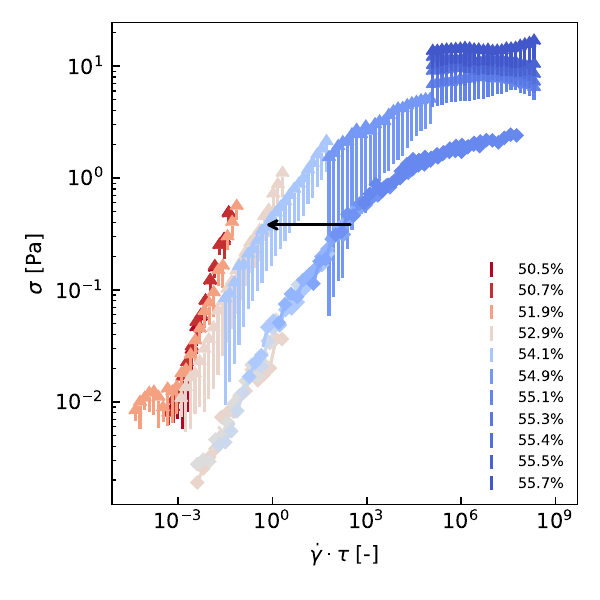}
\caption{
Shear stress $\sigma$ as a function of the scaled shear rate $\dot{\gamma}\tau$. The horizontal shift factors are referenced with the assumption of the same packing fraction of $\varphi=52.9\%$ for CS and SAOS. No vertical scaling is applied. The arrow shows the shift-factor of $9\times10^{-3}$ necessary to collapse the Master Plot.
}
\label{fig7}
\end{figure}

Figure~\ref{fig7} shows the rheological data after applying only the horizontal shift factors required for the time--state superposition (TSS). The characteristic relaxation times are referenced to the packing fraction $\varphi=52.9\%$, while the stress axis remains unscaled.

The successful collapse demonstrates that the dominant effect of changing the fluidization state is a rescaling of the characteristic relaxation time. In close analogy to time--temperature superposition in molecular glass formers, variations in fluidization shift the spectra along the time (or frequency) axis while largely preserving the functional form of the constitutive response.

Systems prepaired with continuous-shear and oscillatory measurements show that identical packing fractions do not necessarily correspond to identical dynamical states. The continuous sheared system is progressively tribocharging and requires a higher fluidization energy to reach the same density. As a result, structurally similar packings exhibit substantially different relaxation dynamics. This demonstrates that packing fraction alone is insufficient to characterize the state of the fluidized bed.

\clearpage

\end{document}